\documentclass[11pt,a4paper]{article}
\usepackage{graphicx}
\usepackage{amsmath,amssymb,graphicx,epsfig}
\begin{document}

\title{Decay of soliton-like perturbations into vortex - anti vortex pairs}

\author{Enik\H{o} J. M. Madarassy$^1$\\
\vspace{1mm}\\
$^1$School of Mathematics, Newcastle University,\\
Newcastle--upon--Tyne, NE1 7RU, UK}


\date{\today}
\maketitle
\begin{center}
(Received \today )
\end{center}

\begin{abstract}
This paper suggests a method to create dark soliton-like waves in a trapped atomic Bose-Einstein condensate (BEC). With the phase imprinting method, a soliton-like disturbance is produced along the $x$-axis. The method is used in two cases (with and without rotation of a trapping potential). The disturbance in a two-dimensional (2D) system is unstable. The soliton-like perturbation is found to decay into more stable vortex pairs together with a burst of sound wave. 
The simulations are performed by solving the 2D Gross-Pitaevskii equation (2D GPE). The successive dynamics of the wave function are studied by monitoring the evolution of density and phase profile. 
\end{abstract}



{\centering\section{Introduction}}
Weakly interacting atomic gases, like $BECs$, give rise to study the nonlinear properties of matter. Soliton solutions of the non-linear Schr\"odinger equation ($NLSE$) first were deduced by Zakharov and Shabat \cite{Zakharov}. Dynamics and stability of dark solitons were described theoretically \cite{Fedichev} together with a method of their creation \cite{Dum}. 
The effectively repulsive atomic interactions and the scattering properties between the atoms of the condensate support the appearance of solitons. 
\bigskip

If the scattering coefficient, $g$ is positive ($g>0$), the effective nonlinearity is {\it repulsive} and describes a {\it dark soliton}. Here, $g=4\pi\hbar^{2}a/m$, $\emph{a}$ is the s-wave scattering length and $m$ is the mass of an atom. 
\bigskip

Dark soliton in matter waves is characterized by a local density minimum and a sharp phase gradient of the wave function at the position of the minimum. Dark solitary wave is not a topological defect like vortices. 
\bigskip

In a dissipative environment, it can disappear anywhere in the condensate. Its quantitative study with a dissipative dynamics at finite temperature was well described \cite{Jackson}. At finite temperature, the dynamics of a vortex within the Zaremba-Nikuni-Griffin formalism was studied in atomic Bose-condensed gas and clarifid its microscopic origin \cite{Jackson1}.
\bigskip

In dilute atomic Bose gas, the quantum theory for the nucleation of Bose-Einstein condensation was presented in \cite{Stoof}. This process can be described by a single Fokker-Planck equation. Symmetry breaking in the self-evolution of strongly non-equilibrium interacting Bose gas can occur only as a result of introduction of a small term of the Hamiltonian or as a result of quantum-mechanical measurements, which imply non-conservation of particles \cite{Kagan}.
\bigskip

Dark solitons are one-dimensional phenomena, which in two- or three-dimensional (2D, 3D) systems are unstable with a transverse instability. Due to the snake-instability \cite{Feder}, they decay into vortex rings \cite{Mamaev}. 
A long-wavelength transverse instability that is analogous to matter-waves was observed \cite{Proukakis} and verified in atomic BECs \cite{Dutton}. 
\bigskip

In harmonic confinement, because of the geometry of the system, a dark soliton-like wave moves with different velocities. 
So, its shape becomes curved, which was also observed experimentally \cite{Denschlag}.
\bigskip

An investigation of fractionalized, half-quantum vortices in $BEC$ was presented in \cite{Ji}. They use sodium atoms and rotating optical traps to study the creation of individual half-quantum vortices and vortex lattices with the help of additional pulsed magnetic trapping potentials. 
\bigskip

They numerically solve the time-dependent coupled $GPE$ of spin-1 and use the same damping rate, $\gamma=0.03$ as we do. Due to the different nature of our $GPE$, our vortices in our presented second case is different than theirs.
\bigskip

Collisions between two orthogonal solitons and their dynamics were studied with coupled $GPE$ in \cite{Zhang}. It was demonstrated that, the exact vector-soliton solution can be obtained with arbitrary tunable interactions. 
Thus, for a sufficiently repulsive interaction in two-component attractive $BEC$, dark-dark solitons can be formed.
\bigskip

Analytical and numerical investigations of dynamics and modulation of ring dark soliton in 2D BEC, were presented in \cite{Hu} with tunable interaction. 
It was shown that, the stability of a shallow ring dark soliton is stable, when the ring is slightly distorted. On the other hand, for large deformation of the ring, vortex pairs appear with different dynamical behaviour. 
\bigskip

In this paper, a phase imprinting method \cite{Burger,Dobrek} is used to study the formation of dark solitary waves in two cases, and the formation of the concurrent acoustic emissions.
\bigskip

{\centering\section{Model}}
To explore the instability of soliton-like waves and the density and phase profiles of a rotating condensate, we solve numerically the following 2D GPE, 
which governs the time evolution of the order parameter, $\Psi(\mathbf{r},t)$:

\begin{equation}
(i-\gamma)\hbar \frac{\partial \Psi}{\partial t} = \left[-\frac{\hbar^{2}}{2m}\nabla_{perp}^{2}
+V_{trap}+g_{2D}N|\Psi|^{2}-\mu-\Omega L_{z}\right]\Psi,
\label{eqn:GP}
\end{equation}
Mean field interactions are represented by $g_\mathrm{2D} = \sqrt{32 \pi} \hbar \omega \frac{a l^{2}}{l_{z}}$ \cite{Regnault}, where $\omega$ is the angular velocity along the $z-axis$, $l$ is the magnetic length, 
$l_{z}$ is the characteristic length of the $z-axis$ oscillator and $\gamma=0.03$ is the dissipation modelling the interaction of the system with the surrounding thermal cloud \cite{Tsubota},\cite{Madarassy}. 
This phenomenological dissipation parameter, $\gamma$, was studied in detail by \cite{Abo-Shaeer}. Its microscopic justification was introduced by Gardiner et al \cite{Gardiner} and Penckwitt et al \cite{Penckwitt}.
The chemical potential, $\mu$, can be initially estimated in dimensional form as:

\begin{equation}
\mu= \omega  \sqrt{\frac{Ngm}{\pi}},
\label{eqn:MuI}
\end{equation}
where, \emph{N} is the number of atoms. The trapping potential becomes:

\begin{equation}
       V_\mathrm{tr}(x,y)=\frac{1}{2}m\omega_{\perp}^{2}\left(x^{2}+y^{2}\right),
\label{eqn:VTR}
 \end{equation}
where, $\omega_{\perp}$ is the radial trap frequency. The quantity $\Omega$ is the angular frequency of rotation about the z-axis and $L_{z}$ = $i\hbar(x\partial_{y}-y\partial_{x})$ is the angular momentum operator.
The density and phase of the condensate together with the superfluid local velocity flow are given by $\rho(\mathbf{r},t)=|\psi(\mathbf{r},t)|^{2}$, $S=\tan^{-1}\frac{\mathrm{Im}(\psi(\mathbf{r},t))}{\mathrm{Re}(\psi(\mathbf{r},t))},$ and ${\bf v }= \hbar/m \nabla S({\bf r},t)$. 
The total energy, $E_\mathrm{tot}$ is defined as:

\begin{equation}
E_\mathrm{tot}=E_\mathrm{kin}+E_\mathrm{int}+ E_q+E_\mathrm{trap},
\label{eqn:TotE}
\end{equation}
where the kinetic energy, $E_\mathrm{kin}$, the internal energy, $E_\mathrm{int}$, the quantum energy, $E_q$, and the trap energy, $E_\mathrm{trap}$ are given respectively by:
\begin{equation}
    E_\mathrm{kin}(t) = \int \frac{\hbar^{2}}{2m} \left( \sqrt{\rho({\bf x}, t)}{\bf v}({\bf x}, t)\right)^{2} d^{2}\mathbf{r},
\label{eqn:KE}
\end{equation}

\begin{equation}
      E_\mathrm{int}(t) = \int g \left(\rho({\bf x}, t)\right)^{2} d^{2}\mathbf{r},
\label{eqn:IE}
\end{equation}

\begin{equation}
      E_q(t) = \int  \frac{\hbar^{2}}{2m}\left(\nabla \sqrt{\rho ({\bf x}, t)}\right)^{2} d^{2}\mathbf{r},
\label{eqn:QE}
\end{equation}

\begin{equation}
      E_\mathrm{trap}(t) = \int \rho({\bf x}, t)V_\mathrm{tr}(\mathbf{x})  d^{2}\mathbf{r}.
\label{eqn:TrE}
\end{equation}
The kinetic energy, $E_{kin}$ is a sum of the sound energy, $E_\mathrm{sound}$, and the vortex energy, $E_\mathrm{vortex}$ \cite{Kobayashi}:

\begin{equation}
E_\mathrm{kin} = E_\mathrm{sound}+E_\mathrm{vortex}.
\label{eqn:KSVE}
\end{equation}
At a given time, $t$, the vortex energy, $E_\mathrm{vortex}$, is achieved by relaxing the $GPE$ in imaginary time, which gives the lowest energy state for the vortex configuration \cite{Parker}. 
Finally, the sound energy is obtained as: $E_\mathrm{sound} = E_\mathrm{kin}-E_\mathrm{vortex}$.
Dimensionless units are also used, with the units of length, time and energy given by: $\sqrt{\hbar/(2m\omega_{\perp})}$, $\omega^{-1}_{\perp}$ and $\hbar\omega_{\perp}$, (harmonic oscillator units, $h.o.u.$). 
The dimensionless form of the $GPE$ is: 

\begin{equation}
\left (i-\gamma\right )\frac{\partial\psi}{\partial t}=\left [-\frac{1}{2}\nabla^{2}+V_{trap}+C|\psi|^{2}-\mu-\Omega L_{z}\right ]\psi,
\label{eqn:2D_GPE}
\end{equation}
where,

\begin{equation}
V_{trap}=\frac{1}{2}\left( x^{2}+y^{2}\right),
\label{eqn:2D_TP}
\end{equation}
and the dimensionless form of the effective coupling constant becomes \cite{Munoz,Munoz1,Munoz2,Madarassy1}:

\begin{equation}
C \equiv \frac{g_{2D} N}{l_{\perp}^{2} \hbar \omega_{\perp}},
\label{eqn:C1}
\end{equation}
and after some transformations, we obtain: 

\begin{equation}
C = 2 \sqrt{2 \pi} \frac{aN}{l_{z}}.
\label{eqn:C2}
\end{equation}
Here, $N$ is the number of atoms, $\omega_{\perp}$ is the radial trap frequency and $l_{z}$ is the characteristic length of the z-axis oscillator.
If, $C=1400$ and $\omega_{z}^{2}$ $\gg$ $C \omega_{\perp}^{2}/ \sqrt{8 \pi}$ $\Rightarrow$ $\omega_{z}^{2}$ $\gg$ $280 \omega_{\perp}^{2}$.
Say, for example $\omega_{z}^{2}$ $\gtrsim$ $2800 \omega_{\perp}^{2}$ $\Rightarrow$ $\omega_{z}$ $\gtrsim$ $53$ $\omega_{\perp}$, where $\omega_{\perp}$ is the radial trap frequency. Throughout this paper, we use $\emph{C}=1400$.
\bigskip

{\centering\section{Results}}
Dark solitary waves are generated with the help of a phase imprinting method, which means a complete absence of fluid (zero density). For the wave function an additional phase factor of $e^{i\Delta \phi}$ is imposed with, $\Delta\phi=\pi$ (a longitudinal phase change of $\pi$). 
Two cases (A and B) are distinguished depending on the presence or absence of the angular frequency of rotation about the $z-axis$, ($\Omega=0$ or $\Omega\ne 0$).
\bigskip

\begin{figure}[p]
      \epsfig{figure=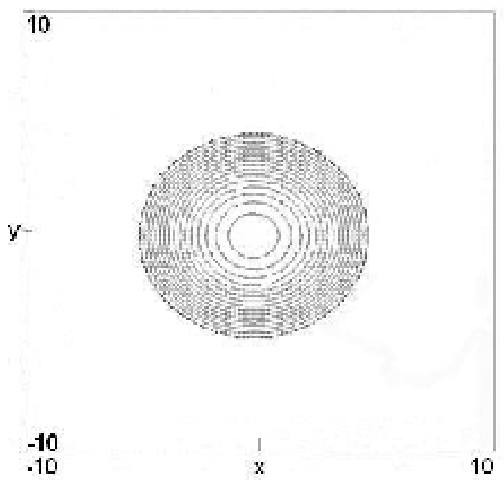,width=0.5\linewidth}
      \epsfig{figure=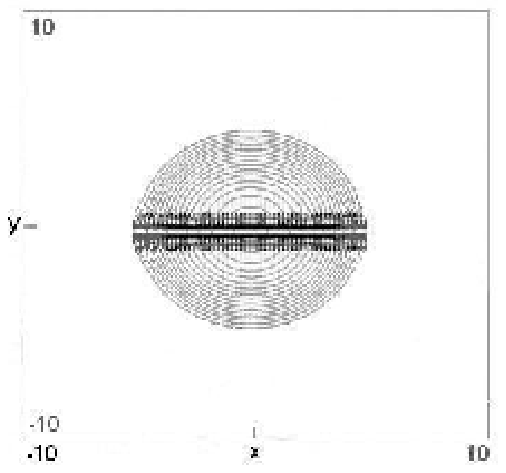,width=0.5\linewidth}
      \caption{Density profile of the condensate for $\Omega=0$. Left) at $t=199$, before the phase imprinting, with no vortices. Right) at $t=200.07$, after the phase imprinting method with soliton-like wave accompanied by large sound waves propagating both in positive and negative directions of the $y$-axis.}
\label{fig:f1}
 \end{figure}

{\centering\subsection{\emph{Phase imprinting in upper two quadrants} for $\Omega=0$.}}
We consider a condensate without vortices ($\Omega=0$, no rotation), as shown in Fig.~\ref{fig:f1} (left). The numerical calculations are performed with the semi-implicit Crank-Nicholson method in a square box of size between $+/-$7 and $+/-$15 (h. o. u.). The origin of the coordinate, $(x,y)=(0,0)$, is at the center of the box. The number of grid points vary between 80 and 150. Over the duration of the run due to the finite discretization of the numerical scheme, the relative change is for example $\Delta E_{tot}/E_{tot}\simeq 0.012$.
\bigskip

\begin{figure}[p]
      \epsfig{figure=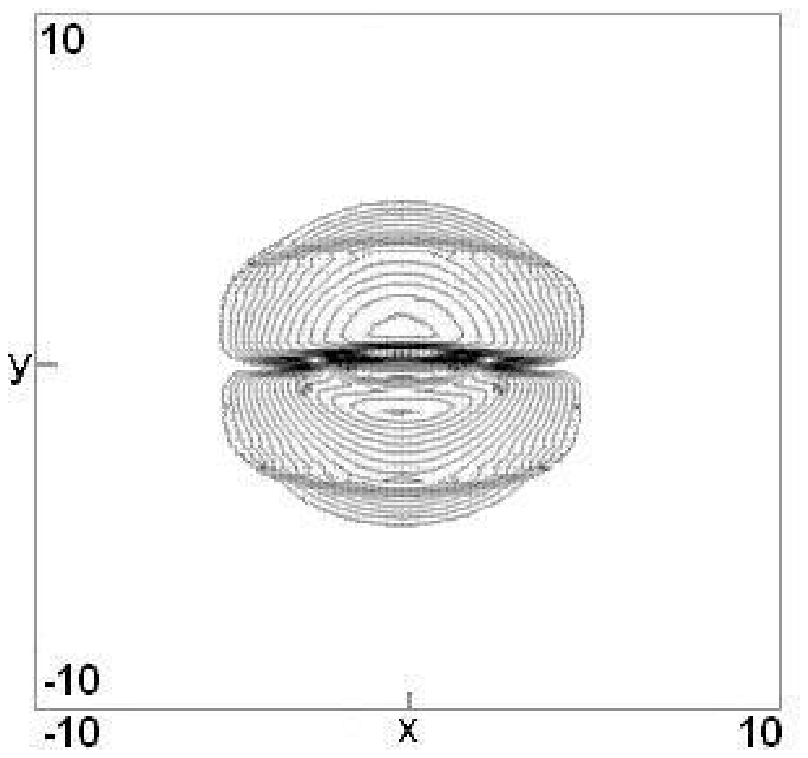,width=0.5\linewidth}
      \epsfig{figure=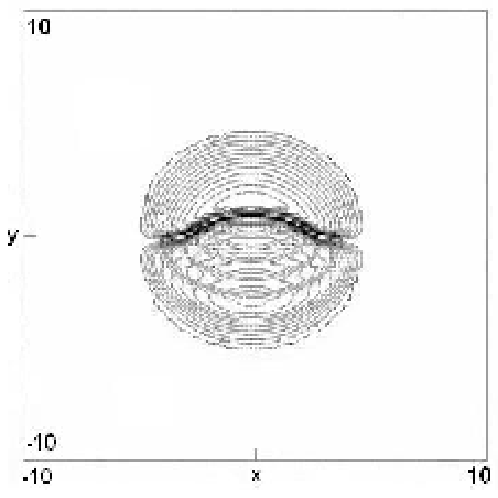,width=0.5\linewidth}
      \caption{Density profile of the condensate corresponding to Fig.~\ref{fig:f1}. Left) at $t=200.45$, the sound waves propagate both upward and downward. Right) at $t=200.6$, a soliton-like perturbation bends and moves, accompanied by sound waves.}
\label{fig:f2}
 \end{figure}

\begin{figure}[p]
      \epsfig{figure=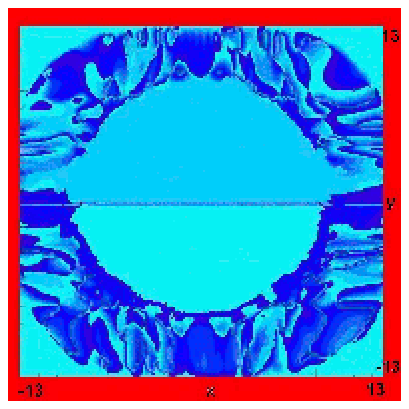,width=0.5\linewidth}
      \epsfig{figure=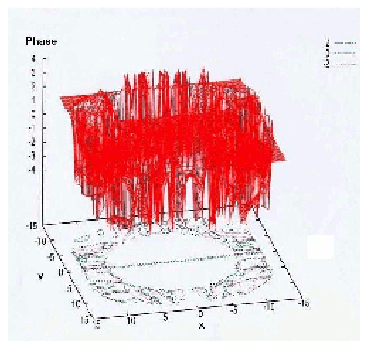,width=0.5\linewidth}
      \caption{Phase profile of the condensate at $t=200$. Left) two-dimensional presentation and Right) three-dimensional presentation. In the random plane region, we have large fluctuations of the phase, (the external part of the condensate), due to $\psi\to 0$, $\mathrm{Im}~\psi\to 0$, $\mathrm{Re}~\psi\to 0$, but the $\mathrm{Phase}=S=\tan^{-1}(\rm{Im}(\psi(\mathbf{r},t))/\rm{Re}(\psi(\mathbf{r},t)))\nrightarrow 0$.}
\label{fig:f4}
 \end{figure}

The \emph{phase} \emph{imprinting} condition is:
 
\begin{equation}
\psi\to\psi~~~\mathrm{for}~~~y>0,
\label{eqn:PIM_IIa}
\end{equation}

and

\begin{equation}
\psi\to e^{i\pi}\psi=-\psi~~~\mathrm{for}~~~y<0,
\label{eqn:PIM_IIb}
\end{equation}
as Fig.~\ref{fig:f4} shows. At $t=200$, $\gamma$ is set to zero. Fig.~\ref{fig:f1} (right) presents the condensate just after the phase imprinting with large sound waves propagating in opposite directions. At $t=200.45$ these sound waves reach almost the edge of the condensate as we can see in Fig.~\ref{fig:f2} (left).
\bigskip

In a confined $BEC$ the density is inhomogeneous. The solitary wave embedded in a $2D$ geometry leads to dominant decay mechanism. Its perturbation has a longitudinal dynamical instability, and it begins to be excited transversely. The applied phase slip of $\pi$ accelerates the soliton-like perturbation due to an interaction with the harmonic trap. 
\bigskip

\begin{figure}[p]
      \epsfig{figure=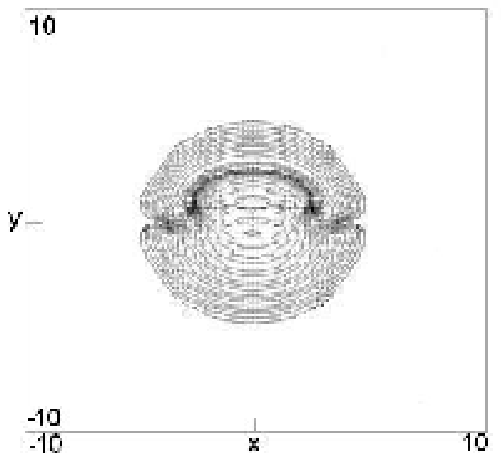,width=0.5\linewidth}
      \epsfig{figure=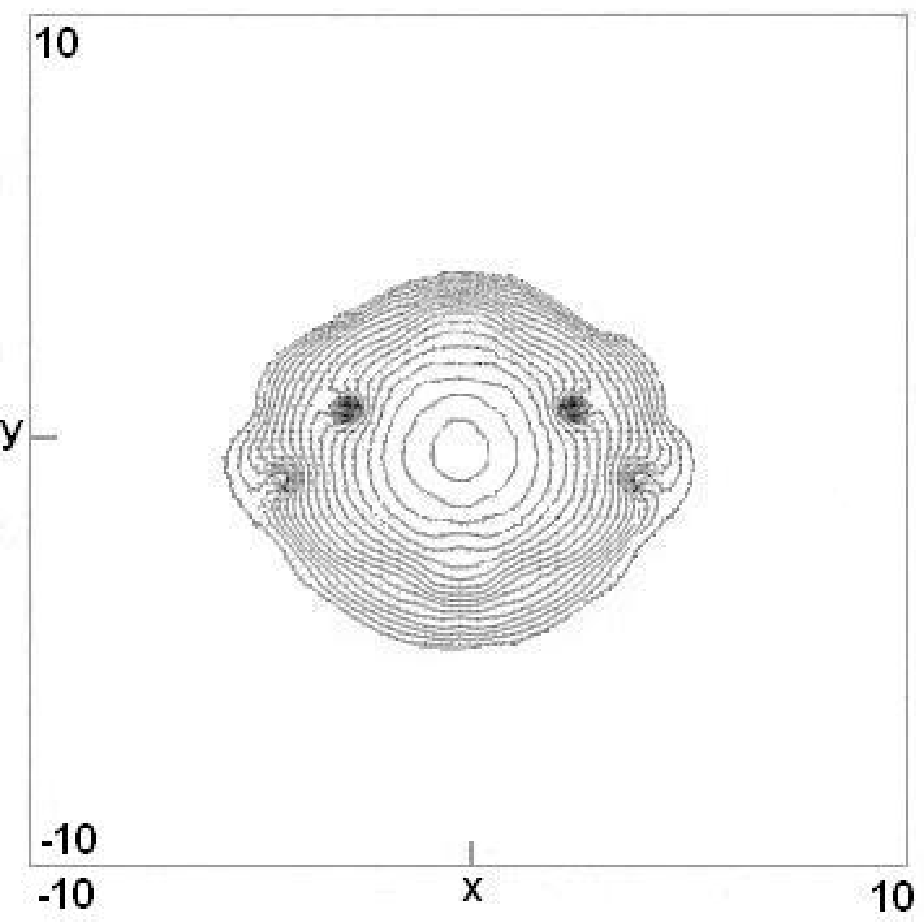,width=0.5\linewidth}
      \caption{Density profile of the condensate corresponding to Fig.~\ref{fig:f1}. Left) at $t=200.75$, the soliton-like perturbation bends more and decays into vortex - anti vortex pairs accompanied by sound waves. Right) at $t=201.15$, the system decays into vortex pairs. The surviving two pairs from the original five pairs of vortices.}
\label{fig:f3}
 \end{figure}

The slightly curved line becomes torn apart and is accompanied by emission of sound waves, which in atomic condensates are density waves. These sound waves propagate in opposite direction to the movement of the solitary wave and are observed in Fig.~\ref{fig:f2} (right) and in Fig.~\ref{fig:f3} (left). 
Close to the center, the solitary wave propagates faster and close to the boundary slower, see Fig.~\ref{fig:f2} (right) and Fig.~\ref{fig:f3} (left). 
\bigskip

A system with solitons tend to decay into more stable, lower energy structures, namely vortex - anti vortex structures in $2D$ or vortex rings in $3D$. Thus, in our case, the soliton-like perturbation bends more and decays into pairs of vortices with opposite charges \cite{Muryshev} via emission of sound waves, see Fig.~\ref{fig:f3} (left).
\bigskip

\begin{figure}[p]
\centering \epsfig{figure=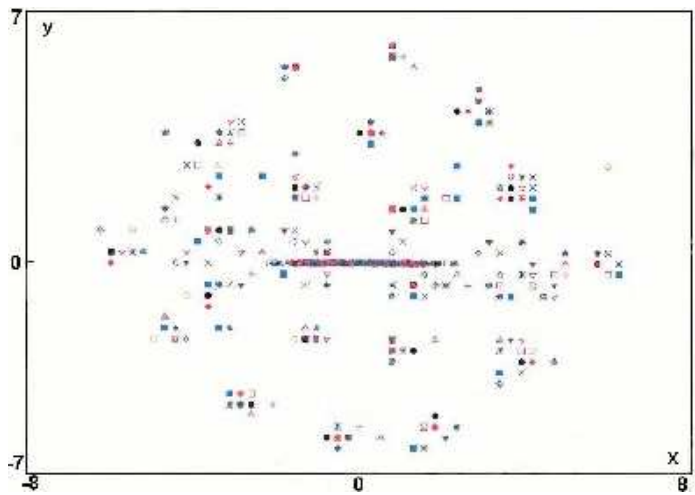,width=0.7\linewidth}
\caption{For $\Omega=0.8$, after applying the phase imprinting method the vortices move in a disordered fashion and the soliton-like perturbation decays into vortex - anti vortex pairs in the center of the condensate, (for $200\le t\le 200.25$).}
\label{fig:f5}
\end{figure}

{\centering\subsection{\emph{Phase imprinting in upper two quadrants} for $\Omega$ = 0.8.}}
In the second case, the system rotates at angular velocity $\Omega$ around the z-axis. Due to a dynamical instability of the condensate, vortex nucleations occur. A stable lattice of 20 vortices is created in a rotating frame. At $t=200$, the previously presented phase imprinting method is suddenly performed and $\gamma$ is set to zero. 
\bigskip
The phase imprinting method has an effect on the other vortices as well. Initially, there are twenty vortices. After this method, four vortex - anti vortex pairs are formed on the line $y=0$. Fig.~\ref{fig:f5} shows the displacement of every vortices in the vortex lattice, and the appearance of vortex - anti vortex pairs in the phase slip due to the soliton like perturbation. 
The corresponding time for these vortex-motions is: $200\le t\le 200.25$.
\bigskip

{\centering\section{Discussion}}
When a discontinuity in the phase is generated, the system tries to smooth out this change and generate dark solitary and sound waves. Thus, after the application of the suggested method solitary wave is formed in a system with a vortex lattice. This method can be applied to create systems with positive and negative vortices in a disc-shaped condensate, whose interactions lead to a disordered motion of vortices. 
\bigskip

\begin{figure}[p]
      \epsfig{figure=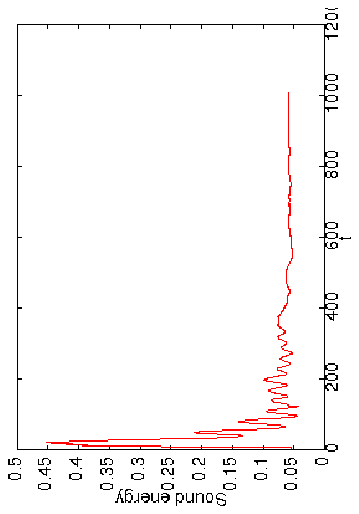,width=0.5\linewidth}
      \epsfig{figure=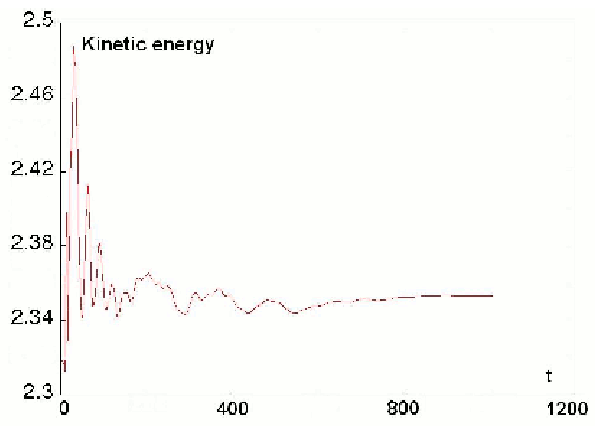,width=0.5\linewidth}
      \caption{For $\Omega=0.8$, Left) Sound energy and Right) Kinetic energy as a function of time, from that time when the phase imprinting method was applied on $y=0$, (here $t=0$ corresponds to $t=200$).}
\label{fig:f6}
 \end{figure}

\begin{figure}[p]
\centering \epsfig{figure=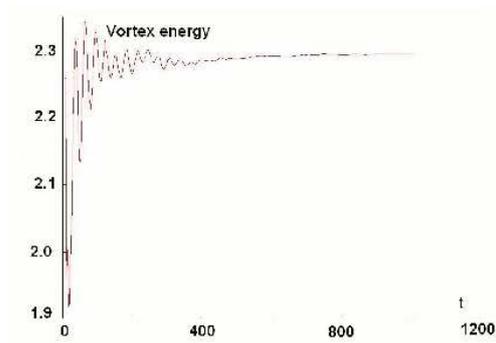,width=0.5\linewidth}
\caption{Vortex energy corresponding to FIG. 6.}
\label{fig:f7}
\end{figure}

Fig.~\ref{fig:f2} (right) and Fig.~\ref{fig:f3} (left) reveal that, large density oscillations (sound waves) are present in the condensate. 
Due to the confined and finite size of the condensate, the sound waves are reflected from the edge of the condensate and interact again with the vortex pairs. We can see the anti-correlation between $E_\mathrm{sound}$ and $E_\mathrm{vortex}$. 
When $E_\mathrm{sound}$ is approaching its maximum, $E_\mathrm{vortex}$ is approaching its minimum, as shown in Fig.~\ref{fig:f6} (left) and Fig.~\ref{fig:f7}. We confirm that $E_\mathrm{sound}=E_\mathrm{kin}-E_\mathrm{vortex}$; compare Fig.~\ref{fig:f6} (left) and (right) with Fig.~\ref{fig:f7}.
\bigskip

\begin{figure}[p]
      \epsfig{figure=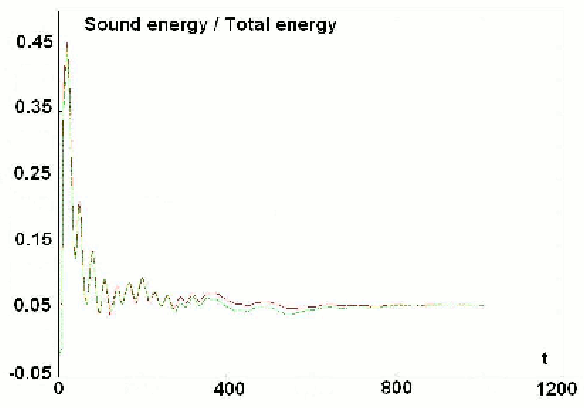,width=0.5\linewidth}
      \epsfig{figure=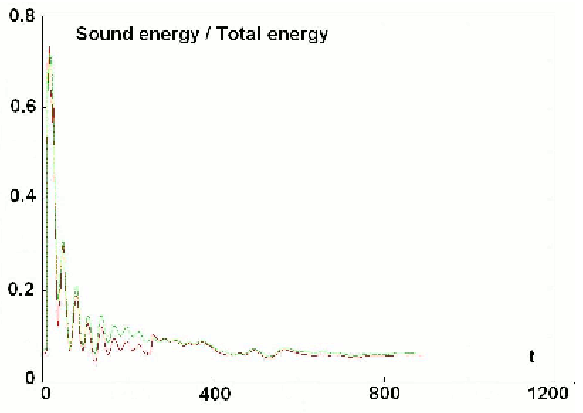,width=0.5\linewidth}
      \caption{For $\Omega=0.8$, the Total energy $-18$, $E_{tot}-18$, is compared with the Sound energy, $E_{sound}$, vs time for Left) the phase imprinting in the upper left quadrant and bottom right quadrant (on both $y=0$ and $x=0$ lines), where the Sound energy is the lower/green curve and for Right) the phase imprinting in the upper two quadrants (on $y=0$), where the Sound energy is the lower/red curve.}
\label{fig:f8}
 \end{figure}

\begin{figure}[p]
      \epsfig{figure=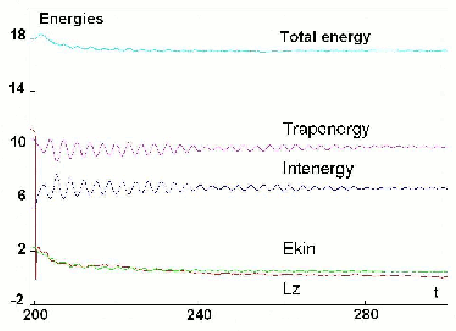,width=0.5\linewidth}
      \epsfig{figure=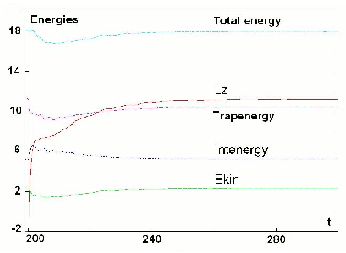,width=0.5\linewidth}
      \caption{ Different energies and $L_{z}$ vs time after the application of the phase imprinting method at $t=200$ for Left) $\Omega=0$ (the anti - vortices are kept inside of the condensate and turbulence is created) and for Right) $\Omega=0.8$ (when the system will relax to the initial vortex lattice). Compare $L_{z}$ and $E_{kin}$ in these two cases.}
\label{fig:f9}
\end{figure}

After applying the phase imprinting method the total energy begins to oscillate and later it relaxes to its final value. We remark that, the sound energy has the biggest contribution to the change of the total energy, which comes from the phase change, from the movement of the vortices and from the interaction between vortices and anti - vortices. 
This is illustrated by the Fig.~\ref{fig:f8}, where the phase imprinting method is applied on both $y=0$ and $x=0$ axis (left) or on $y=0$ axis (right). Here, $E_\mathrm{tot}$ is lowered down by $18$ units to be possible the comparison. 
From Fig.~\ref{fig:f9} (left) and (right), we can see the real value (in h.o.u.) of the total energy. The compressible kinetic energy is directly concerned with the acoustic emission.
\bigskip

For $\Omega=0.8$, after a certain transition period, the system enters the state that is identical to the original regular vortex lattice. This is explained by the fact that although we start with different vortex configurations (no vortices versus some artificial vortex - anti vortex pairs), 
the calculations produce identical results. In that case, the negative vortices will move to the boundary of the condensate. The corresponding variation of different energies and $L_{z}$ are presented by Fig.~\ref{fig:f9} (right). 
\bigskip

To create turbulence with the double use of this method \cite{Madarassy2} on both $y=0$ and $x=0$ axis, immediately after the application of this method, $\Omega$ is set to $0$ at $t=200$, Fig.~\ref{fig:f9} (left) shows the energies and the $z$-component of the angular momentum for that case.
\bigskip

We can compare this method to produce turbulence with a different one, when $\Psi$ = $\Psi^{*}$ on the upper half part of the box $(y>0)$. Here, $\Psi^{*}$ is the complex conjugate of $\Psi(\mathbf{r},t)$ \cite{Madarassy1}. 
In that case, the number of anti - vortices and the shape of the condensate are different due to the fact that more anti - vortices and annihilations appear.
\bigskip

{\centering\section{Acknowledgements}}
The author is very grateful to Carlo F. Barenghi and Vicente Delgado for suggestions and discussions.



\end{document}